\documentclass[letterpaper,10pt,twocolumn]{article}

\usepackage{graphicx}
\usepackage{amsmath}
\usepackage{amssymb}

\newcommand{\dirac}{{\slash \negthinspace \negthinspace \negthinspace \nabla}}
\newcommand{\dd}{\textrm{d}}
\newcommand{\im}{{\mathbb{I}}{\mathrm{m}}}

\title{The Dirac equation in $D$-dimensional spherically symmetric spacetimes}

\author{A. L\'opez-Ortega\thanks{alopezo@ipn.mx} \\
Centro de Investigaci\'on en Ciencia Aplicada y Tecnolog\'{\i}a Avanzada. \\
	      Unidad Legaria. Instituto Polit\'ecnico Nacional. \\
              Calzada Legaria \# 694. Colonia Irrigaci\'on. Delegaci\'on Miguel Hidalgo. \\
	      M\'exico, D.\ F., M\'exico. \\
	      C.\ P.\  11500  
}

\begin{document}

\maketitle

\begin{abstract}

We expound in detail a method frequently used to reduce the Dirac equation in $D$-di\-men\-sion\-al ($D \geq 4$) spherically symmetric spacetimes to a pair of coupled partial differential equations in two variables. As a simple application of these results we exactly calculate the quasinormal frequencies of the uncharged Dirac field propagating in the $D$-di\-men\-sion\-al Nariai spacetime.

\end{abstract}

\section{Introduction}
\label{sect 1}

Recently in many research lines of theoretical physics the models in which the spacetime has more dimensions than the four dimensions observable in our daily experience have been studied extensively. The most analyzed models are those related to string theory \cite{Polchinski-book}. Also the scrutiny of the properties and solutions of higher di\-men\-sion\-al general relativity has attracted a lot of attention (see Ref.\ \cite{Emparan:2008eg} and references therein). In several of these research lines we need to know the classical properties of the higher di\-men\-sion\-al spacetimes to examine different phenomena. Therefore the investigation of these classical properties is an active research field.

To analyze the classical properties of a given spacetime a common method is to use a field as probe \cite{Kokkotas:1999bd}, \cite{Chandrasekhar book}. Thus in the past several scattering phenomena of classical fields were studied, in order to know how to calculate the physical parameters of the spacetime from the measured values of the physical quantities corresponding to the classical field. 

The quasinormal modes (QNMs) are solutions to the equations of motion for a classical field that satisfy the radiation boundary conditions that are natural  in the spacetime in which the field is propagating \cite{Kokkotas:1999bd}, \cite{Chandrasekhar book}. For example, in asymptotically flat black holes the boundary conditions of the QNMs are that the field is purely ingoing near the event horizon and purely outgoing near infinity \cite{Kokkotas:1999bd}. For asymptotically anti-de Sitter black holes we impose the boundary condition that the field vanishes at infinity and is ingoing near the event horizon. 

It has been shown that the QNMs are a useful tool to calculate the physical parameters of a spacetime \cite{Kokkotas:1999bd}, \cite{Chandrasekhar book}. Hence if we know the  quasinormal frequencies (QNF) of a classical field we can infer the values of several physical quantities of the spacetime such as its mass, charge, and angular momentum \cite{Kokkotas:1999bd}. Furthermore it has been proposed that the QNMs encode some information about the quantum properties of the black holes \cite{Hod:1998vk}.

To compute the QNF of a classical field in a given spacetime the usual procedure is to reduce the equations of motion for the field to a radial ordinary differential equation (assuming a given dependence on the angular variables and a harmonic time dependence) and impose to the radial function the boundary conditions of the QNMs. 

Also notice that the reduced form of the equations of motion is useful (and sometimes necessary) to study many other classical or semiclassical phenomena. Thus we believe that at present time the understanding of the separability properties of the  equations of motion for classical fields in higher di\-men\-sion\-al curved spacetimes must be a relevant part in the education of a physicist. 

Motivated by these theories that assume a number of spacetime dimensions greater than four, the separability properties of the equations of motion for several classical fields were studied in higher di\-men\-sion\-al backgrounds. It was found that many of the well known results that are true for four-di\-men\-sion\-al spherically symmetric spacetimes extend to $D$-di\-men\-sion\-al ($D \geq 4$) spherically symmetric spacetimes. 

For example, the reduction of the equations of motion for Klein-Gordon, electromagnetic, and gravitational perturbations to ordinary differential equations which is true in four-di\-men\-sion\-al uncharged spherically symmetric spacetimes \cite{Chandrasekhar book}, also is valid in $D$-di\-men\-sion\-al uncharged spherically symmetric spacetimes, as showed in Refs.\ \cite{Kodama:2003jz}, \cite{Ishibashi:2003jd}. Moreover for the coupled gravitational and electromagnetic perturbations the reduction of the equations of motion to Schr\"odinger type equations which is true in four-di\-men\-sion\-al charged spherically symmetric backgrounds also is valid for $D$-di\-men\-sion\-al charged spherically symmetric backgrounds \cite{Kodama:2003kk}.  

Although the study of the classical dynamics of fields in curved spacetimes (in four and $D$ dimensions) is focused on boson fields \cite{Kokkotas:1999bd}, \cite{Chandrasekhar book}, \cite{Kodama:2003jz}, \cite{Ishibashi:2003jd}, \cite{Kodama:2003kk} mainly on gravitational perturbations, we believe that the understanding of the classical dynamics of the fermion field in $D$-di\-men\-sion\-al spacetimes is of great value, because the Dirac field sometimes behaves in a different way that the boson fields. For example, it is a well known fact that in the four-di\-men\-sion\-al Kerr black hole the fermion field does not show superradiant scattering \cite{Unruh:1974bw}, \cite{Martellini:1977qf}, unlike to boson fields \cite{Chandrasekhar book}. 

For the Dirac equation, its separability properties in $D$-di\-men\-sion\-al spherically symmetric spacetimes were previously studied in Refs.\ \cite{Gibbons:1993hg}, \cite{Cotaescu:1998ay}. In these papers is shown that the Dirac equation reduces to a pair of coupled partial differential equations in two variables.\footnote{Notice that for Dirac field some results valid in four-di\-men\-sion\-al rotating black holes have been extended to rotating black holes in higher dimensions, see Refs.\ \cite{Chandrasekhar:1976ap} for an incomplete list of references. For a review of the recent work on the separability properties for the equations of motion for several fields in higher di\-men\-sion\-al spacetimes see Ref.\ \cite{Kubiznak:2008qp}.} 

Owing to the past and future applications of the reduced system of partial differential equations for the Dirac equation in $D$-di\-men\-sion\-al spherically symmetric spacetimes, we believe that the method used in Refs.\ \cite{Gibbons:1993hg} to reduce the Dirac equation to a pair of coupled partial differential equations deserves a detailed exposition, because this account may be practical and useful. Here we present the method in more detail than in the original references, that is, in the present work we explicitly write some mathematical steps omitted in Refs.\ \cite{Gibbons:1993hg}, \cite{Cotaescu:1998ay} (see also \cite{Gibbons:2008gg}, \cite{Cho:2007zi}, \cite{Cho:2007ce}). 

Notice that Section \ref{sect 2} is not an exhaustive review of the previous work on the dynamics of fermion fields in spherically symmetric spacetimes. Also observe that in this paper we do not consider in detail the mathematical properties of $D$-di\-men\-sion\-al spinors, these can be studied in many books and papers (see for example Refs.\ \cite{Hurley book}, \cite{VanProeyen:1999ni}, \cite{West:1998ey}). We only write the essential properties of the spinors necessary to make the reduction of the Dirac equation to two coupled partial differential equations that we shall expound in Section \ref{sect 2}.

Recently the exact computation of the QNF for several higher and lower di\-men\-sion\-al spacetimes has attracted a lot of attention, see \cite{Cardoso:2001hn}-\cite{Vanzo:2004fy} for some references in which an exact calculation of the QNF was carried out. As many exactly solvable models in theoretical physics, we believe that these examples are useful models and it is possible that they play a relevant role in future research. 

The Nariai spacetime is a vacuum solution to the Einstein equations with positive cosmological constant \cite{b: Nariai solution}. This spacetime is a simple solution to the field equations of general relativity. Owing to this simplicity of the Nariai solution, it is possible to calculate the values of several physical quantities in exact form. For example, for this spacetime in Refs.\ \cite{Vanzo:2004fy} were computed exactly the values of the QNF for Klein-Gordon fields and tensor type gravitational perturbations. 

Furthermore in the $D$-di\-men\-sion\-al charged Nariai spacetime \cite{Kodama:2003kk}, \cite{b: Nariai solution}, the QNF for coupled gravitational and electromagnetic perturbations were calculated exactly in Ref.\ \cite{LopezOrtega:2007vu}. To our knowledge the result of the previous reference in the charged Nariai spacetime is the only exact calculation of QNF for coupled electromagnetic and gravitational perturbations in higher dimensions.

As an application for the reduced system of differential equations obtained in Section \ref{sect 2} for Dirac field moving in $D$-di\-men\-sion\-al spherically symmetric spacetimes, we exactly calculate the QNF of this field in $D$-di\-men\-sion\-al Nariai spacetime \cite{b: Nariai solution}. These values of the QNF for Dirac field extend those already published in Refs.\  \cite{LopezOrtega:2007vu}, \cite{Vanzo:2004fy}. 

In this paper we assume that the reader has a working knowledge of general relativity and differential geometry. Furthermore, in the following sections we use Einstein's sum convention and understand sum on repeated indices (Latin and Greek indices), unless we explicitly state that in a given formula we do not understand sum on repeated indices.

The paper is organized as follows. In Section \ref{sect 2} we present in detail the method of Refs.\ \cite{Gibbons:1993hg} (see also \cite{Gibbons:2008gg}, \cite{Cho:2007zi}, \cite{Cho:2007ce}) that reduces the Dirac equation in $D$-di\-men\-sion\-al spherically symmetric spacetimes to a pair of coupled partial differential equations in two variables. Using these results in Section \ref{sect 3} we exactly calculate the QNF of the Dirac field propagating in the $D$-di\-men\-sion\-al Nariai spacetime. Finally in Section \ref{sect 4} we discuss some related facts.

\section{Dirac's equation in $D$-di\-men\-sion\-al spherically symmetric spacetimes}
\label{sect 2}

As is well known in $D$-di\-men\-sion\-al spherically symmetric backgrounds the Dirac equation 
\begin{equation} \label{eq: Massive Dirac equation}
 i \dirac \psi = m \psi ,
\end{equation} 
reduces to a pair of coupled partial differential equations in two variables \cite{Gibbons:1993hg}, \cite{Cotaescu:1998ay}, \cite{Gibbons:2008gg}, \cite{Cho:2007zi}, \cite{Cho:2007ce}. In this section we describe in detail the method of Refs.\ \cite{Gibbons:1993hg} frequently used to get this result. For a different method see Refs.\ \cite{Cotaescu:1998ay}.

Here we shall consider two $D$-di\-men\-sion\-al spacetimes $M$ and $\tilde{M}$ whose metrics $g_{\mu \nu}$ and $\tilde{g}_{\mu \nu}$ are conformal, that is\footnote{Notice that Greek indices stand for coordinate indices, whereas the Latin indices stand for frame indices.}
\begin{equation} \label{eq: conformal metrics}
 \tilde{g}_{\mu \nu} = \Omega^2  g_{\mu \nu} ,
\end{equation} 
where $\Omega$ is a function of the coordinates. We point out that if the symbols $\dirac$, $\psi$, $m$, and $\tilde{\dirac}$, $\tilde{\psi}$, $\tilde{m}$ denote the Dirac operator, the Dirac spinor, and the mass of the Dirac field corresponding to the spacetimes $M$ and $\tilde{M}$, respectively, then the following relations are satisfied \cite{Gibbons:1993hg}, \cite{Gibbons:2008gg}, \cite{Cho:2007zi}, \cite{Cho:2007ce}
\begin{align} \label{eq: conformal relations}
\psi & = \Omega^{(D-1)/2} \tilde{\psi} , \nonumber \\
\dirac \psi &= \Omega^{(D+1)/2} \tilde{\dirac} \tilde{\psi} , \\
m &= \Omega \tilde{m}. \nonumber
\end{align} 

It is well known that the previous results can be generalized when there are gauge fields \cite{Gibbons:1993hg}, but in this paper we do not analyze this extension. As in formulae (\ref{eq: conformal relations}), in the rest of the present section a tilde stands for quantities corresponding to the spacetime with metric $\tilde{g}_{\mu \nu}$.

One way to obtain the results of formulae (\ref{eq: conformal relations}) is the following. When the metrics $g_{\mu \nu}$ and $\tilde{g}_{\mu \nu}$ are conformally related as in formula (\ref{eq: conformal metrics}) and we define the basis of one-forms $e^a$ such that $g_{\mu \nu} = e_\mu^a e_\nu^b \eta_{a b}$ and the basis $\tilde{e}^a$ such that $\tilde{g}_{\mu \nu} = \tilde{e}_\mu^a \tilde{e}_\nu^b \eta_{a b}$, where $\eta_{a b} = \eta^{ab} = \textrm{diag}(1,-1,\dots,-1)$ is the Minkowski metric \cite{Winitzki lectures}, \cite{Nakahara book}, we find that the one-forms $e^a$ and $\tilde{e}^a$ satisfy
\begin{equation}
 \tilde{e}^a = \Omega \, e^a.
\end{equation} 

To get the relation between the connection one-forms $\omega^a_{\,\,\,b}$ and $\tilde{\omega}^a_{\,\,\,b}$ corresponding to the basis of one-forms $e^a$ and $\tilde{e}^a$, respectively, we recall that the connection one-forms $\omega^a_{\,\,\,b}$ are determined by the first Cartan structure equation 
\begin{equation}
 \dd e^a = - \omega^a_{\,\,\,b} \wedge e^b,
\end{equation} 
where the symbol $\wedge$ stands for the wedge product \cite{Winitzki lectures}, \cite{Nakahara book}. 

Using Statement\footnote{Statement 6.1.6.1 of Ref.\ \cite{Winitzki lectures}: If any set of 2-forms $A_c$ is given and $\{ \theta^c \}$  is a dual frame basis then there exists a unique set of 1-forms $\chi_{ab}$ such that 
\begin{equation}
 A_a + \sum_b \chi_{ab} \wedge \theta^b = 0, \qquad \chi_{ab} = - \chi_{ba} . \nonumber
\end{equation} 
The 1-forms $\chi_{ab}$ can be expressed by the formula
\begin{equation}
 \chi_{ab} = \sum_c \chi_{abc} \theta^c, \qquad \chi_{abc} \equiv \frac{1}{2}(A_{abc} - A_{bac} - A_{cab}), \nonumber
\end{equation}
where $A_{cab}$ are the coefficients in the decomposition
\begin{equation} 
 A_c = \frac{1}{2}\sum_{a,b} A_{cab} \theta^a \wedge \theta^b, \qquad A_{cab} = - A_{cba}. \nonumber
\end{equation}  } 6.1.6.1 of Ref.\ \cite{Winitzki lectures}, we see that the one-forms $\tilde{\omega}^a_{\,\,\,b}$ and $\omega^a_{\,\,\,b}$ are related by 
\begin{equation} \label{eq: dirac equation conformal}
 \tilde{\omega}_{ab} = \omega_{ab} + \frac{1}{\Omega}( e_b(\Omega) e_a - e_a(\Omega) e_b  ),
\end{equation} 
where $e_a(\Omega)$ denotes the action of the vector $e_a = \eta_{ab} e^b$ on the scalar function $\Omega$. From expression (\ref{eq: dirac equation conformal}) we get
\begin{equation}
  \tilde{\omega}_{abc} = \frac{\omega_{abc}}{\Omega} + \frac{1}{\Omega^2}( e_b(\Omega) \eta_{ac} - e_a(\Omega) \eta_{bc} ) ,
\end{equation} 
where $\omega_{abc} = \omega_{ab} (e_c)$ and similarly for $\tilde{\omega}_{abc}$.

Thus if the symbol $\nabla_c$ stands for covariant derivative of a spinor, that is, \cite{Hurley book}, \cite{Nakahara book} 
\begin{equation}
\nabla_c = e_c + \frac{1}{4} \omega_{abc}\gamma^a \gamma^b
\end{equation}
where $\gamma^c$ stands for $D$-di\-men\-sion\-al gamma matrices that satisfy \cite{Hurley book}, \cite{VanProeyen:1999ni},  \cite{West:1998ey}
\begin{equation}
 \gamma^a \gamma^b + \gamma^b \gamma^a = 2 \eta^{a b},
\end{equation} 
and we also observe that $\tilde{\gamma^a} = \gamma^a $. Taking into account the previous results and if we take $\psi = \Omega^s \tilde{\psi}$ then it is possible to show that the $D$-di\-men\-sion\-al Dirac operator 
\begin{equation}
 \dirac \psi = \gamma^c \nabla_c \psi ,
\end{equation} 
transforms into \cite{Gibbons:1993hg}, \cite{Gibbons:2008gg}, \cite{Cho:2007zi}, \cite{Cho:2007ce}
\begin{align} 
 \dirac \psi & = \tilde{\gamma}^c \left[ \Omega \tilde{e}_c +\frac{1}{4}\left( \tilde{\omega}_{abc} \Omega + \tilde{e}_a (\Omega) \eta_{bc} \right. \right.  \nonumber \\
& \hspace{2.5cm}\left. \left. - \tilde{e}_b(\Omega) \eta_{ac} \right) \tilde{\gamma}^a \tilde{\gamma}^b \right] \Omega^s \tilde{\psi}  \\ 
 & = \left[ \tilde{\gamma}^c \Omega \tilde{e}_c + \frac{\Omega}{4} \tilde{\omega}_{abc} \tilde{\gamma}^c \tilde{\gamma}^a \tilde{\gamma}^b - \frac{D-1}{2} \tilde{\gamma}^c \tilde{e}_c (\Omega) \right]  \Omega^s \tilde{\psi} \nonumber \\ 
& = \Omega^{s+1} \tilde{\gamma}^c \tilde{e}_c \tilde{\psi} + \frac{\Omega^{s+1}}{4} \tilde{\omega}_{abc} \tilde{\gamma}^c \tilde{\gamma}^a \tilde{\gamma}^b  \tilde{\psi}   \nonumber \\
&  \hspace{2cm} + \Omega^{s} \tilde{\gamma}^c \tilde{e}_c (\Omega) \left(s - \frac{D-1}{2}\right) \tilde{\psi}. \nonumber
\end{align} 

Taking $s=(D-1)/2$ in the previous formula we finally get the result
\begin{equation} \label{eq: spinor tilde}
 \dirac \psi = \Omega^{(D+1)/2} \tilde{\dirac} \tilde{\psi}.
\end{equation} 
This expression and $\psi = \Omega^{(D-1)/2} \tilde{\psi} $ are the first two results of formulae (\ref{eq: conformal relations}). The result given in expressions (\ref{eq: conformal relations}) for mass $m$ immediately follows from the previous formulae for $ \dirac \psi$ and $\psi$.

In the following paragraphs we study the Dirac equation in the $D$-di\-men\-sion\-al spherically symmetric spacetimes with coordinates $(t,r,\phi_i)$, where $i=1, 2, \dots , (D-2)$, and whose line elements we write in the form
\begin{equation} \label{eq: general metric}
\dd s^2 = F(r)^2 \dd t^2 -G(r)^2 \dd r^2 - H(r)^2 \dd \Sigma^2_{D-2},
\end{equation} 
where $F(r)$, $G(r)$, and $H(r)$ are functions only of the coordinate $r$, the symbol $\dd \Sigma^2_{D-2}$ stands for line element of a $(D-2)$-di\-men\-sion\-al invariant base space which depends only on the coordinates $\phi_i$.

To simplify the Dirac equation in a spacetime whose metric takes the form (\ref{eq: general metric}), we factor out the function $H(r)^2$ in the line element (\ref{eq: general metric}) and define
\begin{equation} \label{eq: H conformal factor}
 \dd \tilde{s}^2 = \frac{ \dd s^2 }{H(r)^2} ,
\end{equation} 
where\footnote{We shall write the functions $F(r)$, $G(r)$, and $H(r)$ simply as $F$, $G$, and $H$, respectively. In general, we shall use a similar convention for the functions that we shall define in the following paragraphs.}
\begin{equation} \label{eq: conformal metric one}
 \dd \tilde{s}^2 = \tilde{g}_{\mu \nu} \dd x^\mu \dd x^\nu = \frac{F^2}{H^2} \dd t^2 - \frac{G^2}{H^2} \dd r^2 -\dd \Sigma^2_{D-2}.
\end{equation} 
Next we use the results (\ref{eq: conformal relations}) to find the relation among the quantities $\psi$, $\dirac \psi$, $m$, and $\tilde{\psi}$, $\tilde{\dirac} \tilde{\psi}$, $\tilde{m}$, corresponding to the spacetimes with line elements $\dd s^2$ of formula (\ref{eq: general metric}) and $\dd \tilde{s}^2$ of expression (\ref{eq: conformal metric one}). We point out that in the line element (\ref{eq: conformal metric one}) the first two terms depend only on the two coordinates $t$, $r$, and the term $\dd \Sigma^2_{D-2}$ depends only on the $(D-2)$ coordinates $\phi_i$. 

As a basis of one-forms for the spacetime with metric (\ref{eq: conformal metric one}) we choose \cite{Gibbons:1993hg}, \cite{Gibbons:2008gg}, \cite{Cho:2007zi}, \cite{Cho:2007ce}
\begin{align} \label{eq: one-form basis}
 \tilde{e}^t &= \tilde{e}^t(r,t) = f_{(t)}(r,t) \dd t + f_{(r)}(r,t) \dd r , \nonumber \\
\tilde{e}^r &= \tilde{e}^r(r,t) = g_{(t)}(r,t) \dd t + g_{(r)}(r,t) \dd r ,  \\
\tilde{e}^i &= \tilde{e}^i(\phi_k) = h^{i}_{j}(\phi_k) \dd \phi^j , \nonumber
\end{align} 
where $i,j,k=1,2,\dots,D-2$. We prefer this basis because many of the connection one-forms are equal to zero, for example, the connection one-forms $\tilde{\omega}_{tj}$ and $\tilde{\omega}_{rj}$.

It is convenient to observe that in $D$ even dimensions the gamma matrices are square matrices of dimension $2^{D/2} \times 2^{D/2}$ whereas in $D$ odd dimensions these are of dimension $2^{(D-1)/2} \times 2^{(D-1)/2}$ \cite{Hurley book}, \cite{VanProeyen:1999ni}, \cite{West:1998ey}. Thus if in the $D$-di\-men\-sion\-al spacetime with metric (\ref{eq: conformal metric one}) we use the representation of the gamma matrices \cite{Gibbons:1993hg},  \cite{Gibbons:2008gg}, \cite{Cho:2007zi}, \cite{Cho:2007ce}, \cite{VanProeyen:1999ni},
\begin{align} \label{eq: Gamma representation}
\tilde{\gamma}_t &= \sigma_1 \otimes  \sigma_0 \otimes  \sigma_0  \otimes  \sigma_0 \dots = \sigma_1  \otimes \mathbb{I}_{2^{(D-2)/2}}, \nonumber \\
\tilde{\gamma}_r &= i \sigma_2 \otimes  \sigma_0 \otimes  \sigma_0  \otimes  \sigma_0 \dots = i \sigma_2 \otimes  \mathbb{I}_{2^{(D-2)/2}}, \nonumber \\
\tilde{\gamma}_1 &= i \sigma_3 \otimes  \sigma_1 \otimes  \sigma_0  \otimes  \sigma_0  \dots = \sigma_3 \otimes \hat{\gamma}_1, \\
\tilde{\gamma}_2 &= i \sigma_3 \otimes  \sigma_2 \otimes  \sigma_0  \otimes  \sigma_0 \dots = \sigma_3 \otimes \hat{\gamma}_2, \nonumber \\
\tilde{\gamma}_3 &= i \sigma_3 \otimes  \sigma_3 \otimes  \sigma_1  \otimes  \sigma_0 \dots = \sigma_3 \otimes \hat{\gamma}_3, \nonumber \\
&\vdots \nonumber \\
\tilde{\gamma}_{D-2} &= \dots = \sigma_3 \otimes \hat{\gamma}_{D-2} , \nonumber
\end{align} 
where $ \mathbb{I}_{2^{(D-2)/2}}$ is the identity matrix of dimension $2^{(D-2)/2} \times 2^{(D-2)/2}$, the symbol $\otimes$ stands for direct product \cite{Hurley book}, \cite{VanProeyen:1999ni}, $\hat{\gamma}_1$, $\hat{\gamma}_2$, \dots, $\hat{\gamma}_{D-2}$ are a representation of the gamma matrices for a $(D-2)$-di\-men\-sion\-al space with signature $(-,\dots,-)$ and 
\begin{align}
\sigma_0 &=  \left( \begin{array}{cc} 
1 & \,\, 0 \\
0 &\,\, 1
\end{array} \right), \qquad
\sigma_1 =  \left( \begin{array}{cc} 
0 & \,\, 1 \\
1 &\,\, 0
\end{array} \right),  
\nonumber \\
\sigma_2 &=  \left( \begin{array}{cc} 
0 & \,\, -i \\
i &\,\, 0
\end{array} \right), \quad
\sigma_3 =  \left( \begin{array}{cc} 
1 & \,\, 0 \\ 
0 &\,\, -1
\end{array} \right),
\end{align} 
that is, $\sigma_1$, $\sigma_2$, and $\sigma_3$ are the Pauli matrices.  

Using the basis of one-forms (\ref{eq: one-form basis}) and the representation for gamma matrices given in formulae (\ref{eq: Gamma representation}), in the spacetime with line element $\dd \tilde{s}^2$ of formula (\ref{eq: conformal metric one}) we find that the $D$-di\-men\-sion\-al Dirac operator $\tilde{\dirac}$ becomes\footnote{In formulae (\ref{eq: Dirac operator two}) there is no sum on the repeated indices $t$, $r$, and $D-1$.} \cite{Gibbons:1993hg}, \cite{Gibbons:2008gg}, \cite{Cho:2007zi}, \cite{Cho:2007ce}
\begin{align} \label{eq: Dirac operator two}
\tilde{\dirac} & = \tilde{\gamma}^t \tilde{\nabla}_t + \tilde{\gamma}^r \tilde{\nabla}_r + \tilde{\gamma}^1 \tilde{\nabla}_1 + \dots + \tilde{\gamma}^{D-2} \tilde{\nabla}_{D-2}  \nonumber  \\
& = (\hat{\gamma}^t \nabla_t^{(2D)} + \hat{\gamma}^r \nabla_r^{(2D)}) \otimes \mathbb{I}_{2^{(D-2)/2}}   \\
& + \sigma_3 \otimes ( \hat{\gamma}^1 \nabla_1^{(D-2)} + \dots + \hat{\gamma}^{D-2} \nabla_{D-2} ^{(D-2)}) \nonumber \\
& =[\dirac_{2D} \otimes \mathbb{I}_{2^{(D-2)/2}} -i\sigma_3\otimes\dirac_{\dd \Sigma}], \nonumber
\end{align} 
where $\nabla^{(2D)}_{t,r}$ and $\nabla^{(D-2)}_{i}$ stand for covariant derivatives of a spinor in two and $(D-2)$ dimensions respectively, $\hat{\gamma}^t$ and $\hat{\gamma}^r$ are a representation of the gamma matrices in two dimensions, $\dirac_{2D}$ is the Dirac operator on the two-di\-men\-sion\-al spacetime whose line element is
\begin{equation} \label{eq: two dimensional metric}
 \dd s_{2D}^2 = \frac{F^2}{H^2}\, \dd t^2 - \frac{G^2}{H^2}\, \dd r^2 ,
\end{equation}  
and $\dirac_{\dd \Sigma}$ is the Dirac operator on the $(D-2)$-di\-men\-sion\-al submanifold with line element $\dd \Sigma^2_{D-2}$ and with signature $(+,\dots,+)$. 

For many relevant spacetimes $\dd \Sigma^2_{D-2}$ is the line element of a $(D-2)$-di\-men\-sion\-al sphere, but this is not the only option, also the quotients of hyperbolic spaces are possible \cite{Aros:2002te}, \cite{Birmingham:2006zx}, \cite{Vanzo:1997gw}.

Taking the spinor $\tilde{\psi}$ of the spacetime with line element (\ref{eq: conformal metric one}) in the form
\begin{equation} \label{eq: spinor two dimensions}
 \tilde{\psi} (r,t,\phi_i) = \psi_{2D}(r,t) \otimes \chi (\phi_i) ,
\end{equation} 
where $\psi_{2D}(r,t)$ is a two-spinor on the spacetime with line element $ \dd s_{2D}^2$ of formula (\ref{eq: two dimensional metric}) and the functions $\chi (\phi_i)$ satisfy
\begin{equation}
 \dirac_{\dd \Sigma} \chi = \kappa \chi,
\end{equation} 
that is, $\chi$ and $\kappa$ denote the eigenfunctions and eigenvalues of the Dirac operator on the manifold with line element $\dd \Sigma^2_{D-2}$ \cite{Camporesi:1995fb}. From formula (\ref{eq: Dirac operator two}) we obtain that the spinor $\tilde{\psi}$ of expression (\ref{eq: spinor two dimensions}) satisfies
\begin{equation}
 \tilde{\dirac} \tilde{\psi} = [\dirac_{2D} - i \sigma_3 \kappa ] \psi_{2D} \otimes \chi.
\end{equation} 

Thus in the $D$-di\-men\-sion\-al spherically symmetric spacetime with line element (\ref{eq: general metric}) the Dirac equation (\ref{eq: Massive Dirac equation}) reduces to
\begin{equation}
\dirac_{2D} \psi_{2D} = (\kappa i \sigma_3 - i m H \mathbb{I}_2) \psi_{2D}. 
\end{equation} 
Next, from the two-di\-men\-sion\-al line element $\dd s^2_{2D}$ of Eq.\ (\ref{eq: two dimensional metric}) we define the line element $\dd \tilde{s}^2_{2D}$ by
\begin{equation} \label{eq: line element tilde}
 \dd s^2_{2D} = \left( \frac{F}{H} \right)^2 \left( \dd t^2 - \frac{G^2}{F^2} \dd r^2 \right) = \left( \frac{F}{H} \right)^2 \dd \tilde{s}^2_{2D},
\end{equation} 
and using the results (\ref{eq: conformal relations}) we obtain that the two-spinor $\tilde{\psi}_{2D}$ of the spacetime with line element $\dd \tilde{s}^2_{2D}$ satisfies the equation
\begin{equation} \label{eq: Dirac equation two dimensions}
 \tilde{\dirac}_{2D} \tilde{\psi}_{2D} =  \frac{F}{H} (i \kappa \sigma_3 - i m H \mathbb{I}_2) \tilde{\psi}_{2D},
\end{equation} 
where $\tilde{\dirac}_{2D}$ stands for Dirac operator on the two-di\-men\-sion\-al spacetime with line element $\dd \tilde{s}^2_{2D}$ defined in formula (\ref{eq: line element tilde}).

Taking the variable $y$ as
\begin{equation}
 \frac{\dd y}{\dd r} = \frac{G}{F},
\end{equation} 
we find that Eq.\ (\ref{eq: Dirac equation two dimensions}) becomes\footnote{Notice that in Eq.\ (\ref{eq: reduced two dimensional Dirac}) there is no sum on the repeated indices $t$ and $y$.}
\begin{equation} \label{eq: reduced two dimensional Dirac}
(\gamma^t \partial_t + \gamma^y \partial_y) \tilde{\psi}_{2D} = \frac{F}{H}(i\kappa \sigma_3 - i m H \mathbb{I}_2)\tilde{\psi}_{2D}, 
\end{equation} 
where $\gamma^t$ and $\gamma^y$ are a representation of the gamma matrices in two spacetime dimensions. 

Here for two-di\-men\-sion\-al gamma matrices $\gamma^t$ and $\gamma^y$ we use the representation \cite{Gibbons:1993hg}, \cite{Gibbons:2008gg}, \cite{Cho:2007zi}, \cite{Cho:2007ce}
\begin{equation} 
 \gamma^t  = \left( \begin{array}{cc} 
0 & \,\, 1 \\
1 & \,\, 0
\end{array} \right),  \qquad \qquad
\gamma^y  = \left( \begin{array}{cc} 
0 & \,\, -1 \\
1 & \,\, 0 
\end{array} \right) ,
\end{equation} 
to find that in the $D$-di\-men\-sion\-al spherically symmetric spacetime with line element (\ref{eq: general metric}) the Dirac equation (\ref{eq: Massive Dirac equation}) reduces to 
\begin{align} \label{eq: Dirac equation general}
 \partial_t  \psi_2 - \frac{F}{G} \partial_r  \psi_2 & =  \left( i \kappa \frac{F}{H} -  i \mu F \right) \psi_1,  \\
 \partial_t \psi_1 + \frac{F}{G} \partial_r \psi_1 & =  - \left( i \kappa \frac{F}{H} +  i \mu F \right) \psi_2,  \nonumber
\end{align} 
where the functions $\psi_1$ and $\psi_2$ are the components of the two-spinor $\tilde{\psi}_{2D}$, that is
\begin{equation} \label{eq: two spinor components} 
 \tilde{\psi}_{2D}=\left( \begin{array}{c} \psi_1 \\ \psi_2 \end{array} \right).
\end{equation} 

Thus we get that in the $D$-di\-men\-sion\-al spherically symmetric spacetime with line element (\ref{eq: general metric}) the Dirac equation (\ref{eq: Massive Dirac equation}) reduces to the pair of coupled partial differential equations in the variables $t$ and $r$ given in Eq.\ (\ref{eq: Dirac equation general}). This system of two coupled partial differential equations in two variables is the main result that we state in this section and it was previously obtained in Refs.\  \cite{Gibbons:1993hg}, \cite{Cotaescu:1998ay}.

It is convenient to notice that in $D$-di\-men\-sion\-al de Sitter spacetime, whose line element in static coordinates takes the form (\ref{eq: general metric}), when we write Eqs.\ (\ref{eq: Dirac equation general}) for this spacetime we obtain Eqs.\ (11) of Ref. \cite{LopezOrtega:2007sr} that we previously get by using the results of Refs.\ \cite{Cotaescu:1998ay} to reduce the Dirac equation.

Although we study spherically symmetric backgrounds in $D \geq 4$ dimensions, we think that the results obtained in this section also are valid in three-di\-men\-sion\-al spacetimes whose metric can take the form (\ref{eq: general metric}). We cannot compare in straightforward way the reduced system of partial differential equations presented in this section with that of Ref.\ \cite{LopezOrtega:2004cq} because in the previous reference a different basis of one-forms was chosen.

\section{Quasinormal modes of the Dirac field in the $D$-di\-men\-sion\-al Nariai spacetime}
\label{sect 3}

As we previously mentioned, the QNF are complex quantities that depend on the physical parameters of a spacetime. Thus if we know the QNF we can infer the values of several physically relevant quantities of the spacetime. For many relevant backgrounds, for example, Schwarzschild and Kerr black holes, it is not possible to calculate the values of their QNF in exact form, we must use approximate or numerical methods \cite{Kokkotas:1999bd}. 

Nevertheless, at present time we know many higher and lower di\-men\-sion\-al spacetimes whose QNF were computed exactly in Refs.\ \cite{Cardoso:2001hn}-\cite{Vanzo:2004fy}. The systems that allow exact computations of some physical parameters have the advantage that we can analyze in more detail their properties and verify in a simple setting some predictions of the physical theories. Hence we believe that these examples deserve a detailed study. Doubtless these models will be useful in future research.

As an elementary application of the coupled system of partial differential equations (\ref{eq: Dirac equation general}) for Dirac field propagating in the $D$-di\-men\-sion\-al spherically symmetric spacetimes that we present in the previous section, here we exactly compute the QNF for this field in the $D$-di\-men\-sion\-al Nariai spacetime which is a simple vacuum solution of the Einstein equations with positive cosmological constant. 

The line element of the $D$-di\-men\-sion\-al Nariai background is \cite{b: Nariai solution} 
\begin{equation} \label{e: metric Nariai}
{\rm d} s^2 = (1 - \sigma r^2)\, {\rm d} t^2 - \frac{ {\rm d} r^2}{(1 - \sigma r^2) } -  a^2 \,{\rm d} \Sigma_{D-2}^2 ,
\end{equation} 
where 
\begin{equation} \label{eq: sigma Nariai} 
 \sigma = (D-1) \Lambda,
\end{equation} 
$\dd \Sigma_{D-2}^2$ is the line element of a $(D-2)$-di\-men\-sion\-al unit sphere, the constant $a^2$ is equal to
\begin{equation} \label{eq: a Nariai}
 a^2  =  \frac{(D-3)}{(D-1)\Lambda},
\end{equation} 
and the constant $\Lambda$ is related to the cosmological constant. If $\sigma > 0$ then the metric (\ref{e: metric Nariai}) has two cosmological horizons at  \cite{b: Nariai solution}
\begin{equation}
 r= \pm \frac{1}{\sqrt{\sigma}}.
\end{equation} 
In the following we assume that the radial coordinate $r \in (-1/\sqrt{\sigma},+1/\sqrt{\sigma})$.

We note that the $D$-di\-men\-sion\-al Nariai spacetime (\ref{e: metric Nariai}) has the following features \cite{b: Nariai solution}: (a) it has a geometry $ S_2 \times \mathbb{S}^{D-2}$, where $S_2$ stands for two-di\-men\-sion\-al de Sitter spacetime and $\mathbb{S}^{D-2}$ denotes the $(D-2)$-di\-men\-sion\-al sphere, (b) it is spherically symmetric, homogeneous, and locally static, (c) it is geodesically complete. Owing to the $D$-di\-men\-sion\-al Nariai spacetime (\ref{e: metric Nariai}) is spherically symmetric, its metric can be written in the form (\ref{eq: general metric}) with the functions $F$, $G$, and $H$ equal to
\begin{equation}
 F=\frac{1}{G}=(1-\sigma r^2)^{1/2}, \qquad H=a.
\end{equation} 

We define the QNMs of the Nariai spacetime as the modes that are purely outgoing near both horizons \cite{LopezOrtega:2007vu}, \cite{Vanzo:2004fy}. We also notice that the results of this section are an extension of those already published in the previous two references for coupled gravitational and electromagnetic perturbations, Klein-Gordon fields and tensor type gravitational perturbations.

To compute the QNF of the uncharged Dirac field that is propagating in the $D$-di\-men\-sion\-al Nariai spacetime (\ref{e: metric Nariai}), we first write in this spacetime the reduced system of partial differential equations (\ref{eq: Dirac equation general}) for Dirac equation in $D$-di\-men\-sion\-al spherically symmetric spacetimes. We get the following system of partial differential equations
\begin{align}\label{eq: Dirac equation Nariai}
\partial_t \psi_2 - (1 - \sigma r^2) \partial_r \psi_2 &= (1 - \sigma r^2 )^{1/2} \left( \frac{i \kappa}{a} - i m \right) \psi_1 ,  \\
\partial_t \psi_1 + (1 - \sigma r^2) \partial_r \psi_1 &= - (1 - \sigma r^2 )^{1/2} \left(\frac{i \kappa}{a} + i m \right) \psi_2 , \nonumber
\end{align}
where $\kappa$ are the eigenvalues of the Dirac operator on the $D$-di\-men\-sion\-al sphere, that is, $\kappa =  \pm i (l + (D-2)/2)$ and $l=0,1,2,\dots$, \cite{Camporesi:1995fb}.

If we take the components $\psi_1$ and $\psi_2$ of the two spinor $\tilde{\psi}_{2D}$ of formula (\ref{eq: two spinor components}) as 
\begin{align} \label{eq: psi 1 2 ansatz}
\psi_1 = R_1(r)\, \textrm{e}^{-i \omega t } ,\nonumber \\
\psi_2 = R_2(r)\, \textrm{e}^{-i \omega t },
\end{align} 
then Eqs.\ (\ref{eq: Dirac equation Nariai}) transforms into the coupled system of ordinary differential equations
\begin{align} \label{eq: radial one Nariai}
(1 - \sigma r^2)\frac{\dd R_2}{\dd r} + i \omega  & R_2  \nonumber \\ 
 &= (1 - \sigma r^2)^{1/2} \left( \frac{iK}{a} + i m \right) R_1, \nonumber \\ 
(1 - \sigma r^2)\frac{\dd R_1}{\dd r} - i \omega & R_1 \\  
&= (1 - \sigma r^2)^{1/2} \left( \frac{iK}{a} - i m \right) R_2, \nonumber
\end{align}
where we define the quantity $K=-\kappa$. Moreover, defining the following quantities $z=\sqrt{\sigma} r$, $\hat{\omega} = \omega/\sqrt{\sigma} $, $\hat{m}=m/\sqrt{\sigma}$, and $\lambda = iK/a \sqrt{\sigma}$, we find that Eqs.\ (\ref{eq: radial one Nariai}) become 
\begin{align} \label{eq: radial two Nariai}
(1 - z^2)\frac{\dd R_2}{\dd z} + i \hat{\omega}  R_2 = (1 - z^2)^{1/2} ( \lambda + i \hat{m} ) R_1, \nonumber \\ 
(1 - z^2)\frac{\dd R_1}{\dd z} - i \hat{\omega}  R_1 = (1 - z^2)^{1/2} ( \lambda - i \hat{m} ) R_2.
\end{align}
Also notice that $z \in (-1,1)$.

Next, we define (as in Chandrasekhar book's \cite{Chandrasekhar book})
\begin{equation} \label{eq: theta Nariai}
 \theta = \arctan \left( \frac{\hat{m}}{\lambda}\right),
\end{equation} 
that is 
\begin{align}
 \lambda & = \sqrt{\lambda^2 + \hat{m}^2} \cos (\theta) , \nonumber \\ 
\hat{m} & = \sqrt{\lambda^2 + \hat{m}^2} \sin (\theta) ,
\end{align} 
and taking
\begin{align}  \label{eq: theta R1 R2}
R_1 & = \textrm{e}^{-i\theta/2 } \tilde{R}_1, \nonumber \\
R_2 & = \textrm{e}^{i\theta/2 } \tilde{R}_2,
\end{align} 
we see that Eqs.\ (\ref{eq: radial two Nariai}) reduce to 
\begin{align} \label{eq: radial three Nariai}
(1 - z^2)\frac{\dd \tilde{R}_2}{\dd z} + i \hat{\omega} \tilde{R}_2 = (1 - z^2)^{1/2} \alpha_N \tilde{R}_1, \nonumber \\ 
(1 - z^2)\frac{\dd \tilde{R}_1}{\dd z} - i \hat{\omega} \tilde{R}_1 = (1 - z^2)^{1/2}  \alpha_N \tilde{R}_2,
\end{align}
where $\alpha_N = \sqrt{\lambda^2 + \hat{m}^2}$.

From the previous equations we get that the decoupled ordinary differential equations for the functions $\tilde{R}_1$ and $\tilde{R}_2$ are equal to
\begin{align} \label{eq: radial decoupled Nariai}
 \frac{\dd^2 \tilde{R}_1}{\dd z^2} - \frac{z}{1-z^2} \frac{\dd \tilde{R}_1}{\dd z} + \frac{(\hat{\omega}^2 - i \hat{\omega} z) \tilde{R}_1 }{(1-z^2)^2} - \frac{\alpha_N^2 \tilde{R}_1 }{1-z^2} = 0,  \nonumber \\
\frac{\dd^2 \tilde{R}_2}{\dd z^2} - \frac{z}{1-z^2} \frac{\dd \tilde{R}_2}{\dd z} + \frac{(\hat{\omega}^2 + i \hat{\omega} z) \tilde{R}_2}{(1-z^2)^2} - \frac{\alpha_N^2 \tilde{R}_2}{1-z^2}  = 0.
\end{align} 

To solve Eqs.\ (\ref{eq: radial decoupled Nariai}) we make the change of variable
\begin{equation} \label{eq: y Nariai}
 y= \frac{1}{2}(z+1), 
\end{equation} 
with $y \in (0,1)$ and the ansatz
\begin{align}
\tilde{R}_1 = (1-y)^{B_1} y^{C_1} S_1(y) ,  \nonumber \\
\tilde{R}_2 = (1-y)^{B_2} y^{C_2} S_2(y), 
\end{align} 
where
\begin{align}
 B_1 = & \left\{ \begin{array}{l} \frac{i \hat{\omega}}{2} + \frac{1}{2}  \\ \\ - \frac{i \hat{\omega}}{2} \end{array}\right., \nonumber \\
C_1 = & \left\{ \begin{array}{l} - \frac{i \hat{\omega}}{2} + \frac{1}{2}  \\ \\  \frac{i \hat{\omega}}{2} \end{array}\right., \nonumber \\
B_2 = & \left\{ \begin{array}{l} \frac{i \hat{\omega}}{2}  \\ \\  - \frac{i \hat{\omega}}{2} + \frac{1}{2}  \end{array}\right. , \\
C_2 =  & \left\{ \begin{array}{l} - \frac{i \hat{\omega}}{2}  \\ \\   \frac{i \hat{\omega}}{2} + \frac{1}{2}  \end{array}\right. ,\nonumber 
\end{align}
to find that the functions $S_1(y)$ and $S_2(y)$ must be solutions of the hypergeometric differential equation \cite{b:DE-books}
\begin{equation} \label{e: hypergeometric differential equation}
y(1-y) \frac{\dd^2 f}{\dd y^2} + (c - (a +b + 1)y)\frac{\dd f}{{\rm d}y} - a b f   = 0,
\end{equation} 
with parameters (the lower indices $1$ or $2$ determine if the parameter correspond to the function $S_1(y)$ or to the function $S_2(y)$)
\begin{align} \label{eq: constants hypergeometric Nariai}
a_1 & = B_1 + C_1 + i \alpha_N,    \nonumber \\
b_1 & = B_1 + C_1 -i \alpha_N,      \nonumber \\
c_1 & = 2 C_1 +\tfrac{1}{2},      \\
a_2 & = B_2 + C_2 + i \alpha_N, \nonumber \\
b_2 & = B_2 + C_2 -i \alpha_N, \nonumber \\
c_2 & = 2 C_2 +\tfrac{1}{2}  . \nonumber 
\end{align}

In the following we study the function $R_1$ (we obtain similar results for the function $R_2$). Also we take the quantities $B_1$ and $C_1$ as  $B_1= - i\hat{\omega}/2$ and $C_1= i\hat{\omega}/2$. From the previous results we see that if the parameter $c_1$ is not an integer, then we write the function $R_1$ as  \cite{b:DE-books}
\begin{align} \label{eq: R1 Dirac Nariai spacetime}
  & R_1  = \textrm{e}^{-i \theta/2}  (1-y)^{- i\hat{\omega}/2}  \left[ \mathbb{D}_1 \, y^{i\hat{\omega}/2} {}_{2}F_{1}(a_1,b_1;c_1;y) \right.   \\
&\left.  + \mathbb{E}_1 \, y^{1/2-i\hat{\omega}/2} {}_{2}F_{1}(a_1-c_1+1,b_1-c_1+1;2-c_1;y) \right],\nonumber
\end{align} 
where $\mathbb{D}_1$ and $\mathbb{E}_1$ are constants.

At this point we note that the tortoise coordinate for the $D$-di\-men\-sion\-al Nariai spacetime is equal to \cite{LopezOrtega:2007vu}, \cite{Vanzo:2004fy}, \cite{b: Nariai solution}
\begin{equation}
 x = \int \frac{\dd r}{1-\sigma r^2} = \frac{1}{\sqrt{\sigma}} \textrm{arctanh} (z),
\end{equation} 
where $x \in (-\infty, +\infty)$ and from expression (\ref{eq: y Nariai}) we get
\begin{align} \label{eq: Nariai relations coordinates}
 \textrm{as } \quad & x \to - \infty,  & y \approx \, &  \textrm{e}^{2 \sqrt{\sigma} x},  \qquad \,\,\, \textrm{and}  \nonumber \\
\textrm{as } \quad & x \to + \infty,  &  1-y \approx \, & \textrm{e}^{- 2 \sqrt{\sigma} x} .
\end{align}

Thus near the horizon at $r=-1/\sqrt{\sigma}$ (that is as $x \to - \infty$) the function $R_1$ (\ref{eq: R1 Dirac Nariai spacetime}) behaves as
\begin{equation} \label{eq: R1 approximate}
 R_1 \approx \mathbb{D}_1 \textrm{e}^{i \omega x} + \mathbb{E}_1 \textrm{e}^{- i \omega x} \textrm{e}^{\sqrt{\sigma} x}.
\end{equation} 
In the $D$-di\-men\-sion\-al Nariai spacetime to satisfy the QNMs boundary condition near $r=-1/\sqrt{\sigma}$, that is, the function $R_1$ behaves as $\exp{(-i \omega x)}$ if the tortoise coordinate goes to minus infinity $x \to - \infty$, we take $\mathbb{D}_1 = 0$ in formula (\ref{eq: R1 approximate}) and therefore the function $R_1$ becomes
\begin{align} \label{eq: radial Nariai R1}
 R_1 & = \mathbb{E}_1 \textrm{e}^{-i \theta/2} (1-y)^{-i\hat{\omega}/2} y^{1/2 - i \hat{\omega }/ 2} \nonumber \\ 
& \times {}_{2}F_{1}(a_1-c_1+1,b_1-c_1+1;2-c_1;y)  \\
& = \mathbb{E}_1 \textrm{e}^{-i \theta/2} (1-y)^{- i\hat{\omega}/2} y^{1/2-i\hat{\omega}/2}\, {}_{2}F_{1}(\alpha_1,\beta_1;\gamma_1;y). \nonumber
\end{align}

We recall that if the quantity $\gamma-\alpha-\beta$ is not an integer then the hypergeometric function ${}_{2}F_{1}(\alpha,\beta;\gamma;u)$ satisfies \cite{b:DE-books}
\begin{align} \label{e: hypergeometric property z 1-z}
{}_2F_1(\alpha,\beta;\gamma;&u) = \frac{\Gamma(\gamma) \Gamma(\gamma-\alpha-\beta)}{\Gamma(\gamma-\alpha) \Gamma(\gamma - \beta)} \nonumber \\ 
& \times {}_2 F_1 (\alpha,\beta;\alpha + \beta + 1 - \gamma;1-u)  \\
& + \frac{\Gamma(\gamma) \Gamma( \alpha + \beta - \gamma)}{\Gamma(\alpha) \Gamma(\beta)} (1-u)^{\gamma-\alpha -\beta} \nonumber \\ 
& \times {}_2F_1(\gamma-\alpha, \gamma-\beta; \gamma + 1 - \alpha -\beta; 1 -u). \nonumber
\end{align}

Thus if the quantity $\gamma_1 - \alpha_1 - \beta_1$ is not an integer, then using formula (\ref{e: hypergeometric property z 1-z}) we write the radial function (\ref{eq: radial Nariai R1}) as
\begin{align} \label{eq: Nariai f1 1-u} 
 R_1 &= \mathbb{E}_1 \textrm{e}^{-i \theta/ 2}  y^{1/2 - i \hat{\omega}/2} \left[ \frac{\Gamma(\gamma_1) \Gamma(\gamma_1 - \alpha_1 - \beta_1)}{\Gamma(\gamma_1 - \alpha_1) \Gamma(\gamma_1 - \beta_1)}  \right. \nonumber \\
& \times (1-y)^{-i\hat{\omega}/2} {}_2 F_1 (\alpha_1,\beta_1;\alpha_1+\beta_1+1-\gamma_1;1-y) \nonumber \\ 
& +  \frac{\Gamma(\gamma_1) \Gamma( \alpha_1 + \beta_1 - \gamma_1)}{\Gamma(\alpha_1) \Gamma(\beta_1)} (1-y)^{1/2 + i \hat{\omega}/2} \\ 
& \left. \times  {}_2F_1(\gamma_1-\alpha_1, \gamma_1-\beta_1; \gamma_1 + 1 -\alpha_1 - \beta_1; 1 -y) \right] . \nonumber 
\end{align} 
Therefore as $x \to + \infty $, taking into account expressions (\ref{eq: Nariai relations coordinates}), we see that the function  $R_1$ is approximately equal to
\begin{align}  \label{eq: asymptotic Nariai}
 R_1 &\approx  \frac{\Gamma(\gamma_1) \Gamma(\gamma_1 - \alpha_1 - \beta_1)}{\Gamma(\gamma_1 - \alpha_1) \Gamma(\gamma_1 - \beta_1)} \textrm{e}^{i \omega x}  \nonumber \\
& + \frac{\Gamma(\gamma_1) \Gamma( \alpha_1 + \beta_1 - \gamma_1)}{\Gamma(\alpha_1) \Gamma(\beta_1)} \textrm{e}^{-i \omega x} \textrm{e}^{-\sqrt{\sigma} x} .
\end{align}

The boundary condition for QNMs of the $D$-di\-men\-sion\-al Nariai spacetime imposes that the function $R_1$ behaves in the form $\exp{(i \omega x)}$ as $x \to + \infty $. Thus to satisfy the boundary condition of the QNMs for Nariai spacetime near the horizon at $r=+1/\sqrt{\sigma}$ we must cancel the second term in formula (\ref{eq: asymptotic Nariai}). One way is to exploit the zeros of the terms $1/\Gamma(x)$ which are located at $x=-n$, $n=0,1,2,\dots$. Hence to satisfy the boundary condition near the horizon at $r=+1/\sqrt{\sigma}$ we must impose the condition
\begin{equation}
 \alpha_1 = -n, \qquad \textrm{or} \qquad \beta_1 = -n,
\end{equation}  
which imply that the QNF of the Dirac field in $D$-di\-men\-sion\-al Nariai spacetime are determined by the expression
\begin{align} \label{eq: Nariai QN frequencies}
 \hat{\omega} & = \pm \alpha_N - i \left( n + \frac{1}{2} \right).
\end{align} 
A similar computation for radial function $R_2$ also yields the QNF of formula (\ref{eq: Nariai QN frequencies}).

Taking into account that 
\begin{equation}
 \alpha_N = \sqrt{\lambda^2 + \hat{m}^2} = \sqrt{\hat{m}^2 + \frac{(2l + D - 2)^2}{4a^2\sigma}} ,
\end{equation}  
we find that in the $D$-di\-men\-sion\-al Nariai spacetime (\ref{e: metric Nariai}) the QNF of the Dirac field are equal to
\begin{equation} \label{eq: QN frequencies Nariai}
 \omega = \pm \sqrt{\sigma} \sqrt{\hat{m}^2 + \frac{(2l + D - 2)^2}{4a^2\sigma}} - i \sqrt{\sigma} \left( n + \frac{1}{2} \right).
\end{equation} 

In our notation the previously calculated QNF for Klein-Gordon field and tensor type gravitational perturbation are written as \cite{Vanzo:2004fy}
\begin{equation} \label{eq: QNF Nariai Klein-Gordon}
 \omega_{KG} = \pm \frac{\left(l(l+D-3) - \tfrac{D-3}{4}\right)^{1/2}}{a} - i \sqrt{\sigma} \left( n + \frac{1}{2} \right) .
\end{equation} 
We point out that for $l=0$ the QNF (\ref{eq: QNF Nariai Klein-Gordon}) of the Klein-Gordon field and tensor type gravitational perturbation are purely imaginary. This fact was not noted in Ref.\ \cite{Vanzo:2004fy}.

When in the $D$-di\-men\-sion\-al charged Nariai spacetime studied in Ref.\ \cite{LopezOrtega:2007vu} we take the electric charge of the spacetime equal to zero, for electromagnetic and gravitational perturbations of vector type we get their QNF are equal to
\begin{align} \label{eq: QNF Nariai vector}
 \omega_V^+ & = \pm \frac{\left(l(l+D-3) + \tfrac{3D-13}{4}\right)^{1/2}}{a} - i \sqrt{\sigma} \left( n + \frac{1}{2} \right), \nonumber \\
\omega_V^- & = \pm \frac{\left(l(l+D-3) - \tfrac{5D-11}{4} \right)^{1/2}}{a} - i \sqrt{\sigma}\left( n + \frac{1}{2} \right),
\end{align} 
whereas for electromagnetic and gravitational perturbations of scalar type their QNF are
\begin{align} \label{eq: QNF Nariai scalar}
 \omega_S^+ & = \pm \frac{\left(l(l+D-3) - \tfrac{D-3}{4}\right)^{1/2}}{a} - i \sqrt{\sigma} \left( n + \frac{1}{2} \right), \nonumber \\
\omega_S^- & = \pm \frac{(l(l+D-3) - \tfrac{9(D-3)}{4})^{1/2}}{a} - i \sqrt{\sigma} \left( n + \frac{1}{2} \right).
\end{align} 

Notice that in Refs.\ \cite{LopezOrtega:2007vu}, \cite{Vanzo:2004fy}, to calculate the QNF (\ref{eq: QNF Nariai Klein-Gordon}), (\ref{eq: QNF Nariai vector}), and (\ref{eq: QNF Nariai scalar}) of the $D$-di\-men\-sion\-al Nariai spacetime the result for QNF of P\"oschl-Teller potential was used. The QNF of this potential were previously computed in the paper by Ferrari and Mashhoon \cite{Ferrari:1984}. 

It is convenient to observe that for identical values of $D$, $n$, and $\Lambda$ the imaginary part of the QNF (\ref{eq: QN frequencies Nariai}) for Dirac field is identical to the imaginary part of the QNF for gravitational and electromagnetic perturbations  of vector type and scalar type (formulae (\ref{eq: QNF Nariai vector}) and (\ref{eq: QNF Nariai scalar})), and for Klein-Gordon fields and tensor type gravitational perturbations (formula (\ref{eq: QNF Nariai Klein-Gordon})) already computed in Refs.\ \cite{LopezOrtega:2007vu}, \cite{Vanzo:2004fy}. Thus the decay time $\tau=1/\im(\omega)$ is the same for fermion and boson fields studied here and in Refs.\ \cite{LopezOrtega:2007vu}, \cite{Vanzo:2004fy}. Also we point out that for QNMs of these fields the decay time does not depend on the angular momentum number $l$.

For Dirac field the real part of the QNF (\ref{eq: QN frequencies Nariai}) show some differences with respect to real part of QNF for Klein-Gordon, electromagnetic, and gravitational perturbations (\ref{eq: QNF Nariai Klein-Gordon}), (\ref{eq: QNF Nariai vector}), and (\ref{eq: QNF Nariai scalar}). We think that the source of these differences is that in the present section we study the Dirac field, whereas in Refs.\ \cite{LopezOrtega:2007vu}, \cite{Vanzo:2004fy} the fields studied are massless boson fields. 

From formula (\ref{eq: QN frequencies Nariai}), we obtain that for massless Dirac field (Weyl field) its QNF are equal to
\begin{equation} \label{eq: QNF massless Dirac}
 \omega = \frac{2l+D-2 }{2a } - i \sqrt{\sigma} \left( n + \frac{1}{2} \right) .
\end{equation} 
Even in this case the real part of the QNF (\ref{eq: QNF massless Dirac}) shows a different dependence on the angular momentum number that the real part of the QNF for Klein-Gordon, electromagnetic, and gravitational perturbations. The imaginary part is equal to that of the Dirac field. Thus for Dirac field in the $D$-di\-men\-sion\-al Nariai spacetime the decay times of its QNMs do not depend on the mass.

Taking into account the results for QNF (\ref{eq: QN frequencies Nariai}), (\ref{eq: QNF Nariai Klein-Gordon}), (\ref{eq: QNF Nariai vector}), (\ref{eq: QNF Nariai scalar}), and (\ref{eq: QNF massless Dirac}) of the $D$-di\-men\-sion\-al Nariai spacetime, we note that their real and imaginary parts show an explicit dependence on the dimension $D$ of the spacetime (the imaginary part through the parameter $\sigma$ of formula (\ref{eq: sigma Nariai})). For all the fields whose QNF have been calculated and for a given mode number $n$ we infer that the decay time decreases as the dimension of the spacetime $D$ increases. The dependence of the oscillation frequency on the spacetime dimension $D$ is more complicated, it first decreases and then increases as $D$ increases. Moreover for $l\geq3$ and for the same values of $D$, $l$, and $n$ the oscillation frequencies of the massless Dirac field (\ref{eq: QNF massless Dirac}) are greater than the oscillation frequencies for Klein-Gordon, electromagnetic, and gravitational perturbations of formulae (\ref{eq: QNF Nariai Klein-Gordon}), (\ref{eq: QNF Nariai vector}), and (\ref{eq: QNF Nariai scalar}).

As the harmonic time dependence is of the form $\exp(- i \omega t)$ (see formulae (\ref{eq: psi 1 2 ansatz})), in order to have stable QNMs we need that $\im (\omega) < 0$.  We notice that for QNF of formula (\ref{eq: QN frequencies Nariai}) $\im (\omega) < 0$, thus the QNMs of the Dirac field decay in time. Also for Klein-Gordon, electromagnetic, and gravitational perturbations a similar result is true (see formulae (\ref{eq: QNF Nariai Klein-Gordon}), (\ref{eq: QNF Nariai vector}), and (\ref{eq: QNF Nariai scalar})). Therefore we assert that the $D$-di\-men\-sion\-al Nariai spacetime is perturbatively stable under the propagation of these classical fields.

To finish this section, we note that in Refs.\ \cite{LopezOrtega:2007vu}, \cite{Vanzo:2004fy} is shown that the radial differential equations for Klein-Gordon fields, tensor type gravitational perturbations, and coupled electromagnetic and gravitational perturbations reduce to Schr\"odinger type equations with a P\"oschl-Teller potential of the form
\begin{equation} \label{eq: potential Poschl-Teller}
 V(x)=\frac{U}{\cosh^2(\sqrt{\sigma} x)},
\end{equation} 
where the value of the constant $U$ depends on the perturbation type (see Refs.\ \cite{LopezOrtega:2007vu}, \cite{Vanzo:2004fy}). 

Following the procedure of Chapter 10 in Ref.\ \cite{Chandrasekhar book} we transform Eqs.\ (\ref{eq: radial one Nariai}) into a pair of Schr\"odinger type equations with potentials equal to
\begin{align} \label{eq: Morse potential}
 V_{\pm}(x) & = \frac{\left( \frac{(2l+D-2)^2}{4 a^2 \sigma} + \hat{m}^2  \right) \sigma}{\cosh^2(\sqrt{\sigma} x)}  \\
& \pm \frac{\left( \frac{(2l+D-2)^2}{4 a^2 \sigma} + \hat{m}^2 \right)^{1/2} \sigma \sinh (\sqrt{\sigma} x)  }{\cosh^2(\sqrt{\sigma} x)}. \nonumber
\end{align} 
We observe that these potentials are of Morse type (see Table I of Ref.\ \cite{Dutt:1988susy}). 

For plots of the potentials (\ref{eq: potential Poschl-Teller}) and (\ref{eq: Morse potential}) see Figures 1 and 2. We note that for identical values of $D$, $\Lambda$, and $l$  the shape of the potentials (\ref{eq: potential Poschl-Teller}) and (\ref{eq: Morse potential}) is similar, only observe that the height of P\"oschl-Teller potential is smaller than the height of Morse potentials. Notice that in Figure 1 is plotted the P\"oschl-Teller potential corresponding to the QNF $\omega_V^+$, but for P\"oschl-Teller potentials (\ref{eq: potential Poschl-Teller}) corresponding to the QNF $\omega_V^-$, $\omega_S^+$, $\omega_S^-$, and $\omega_{KG}$ also is true that for identical values of $D$, $\Lambda$, and $l$ their height is smaller than the height of Morse potential (\ref{eq: Morse potential}).

\begin{figure}[th]
\label{figure1}
\includegraphics[scale=.85,clip=true]{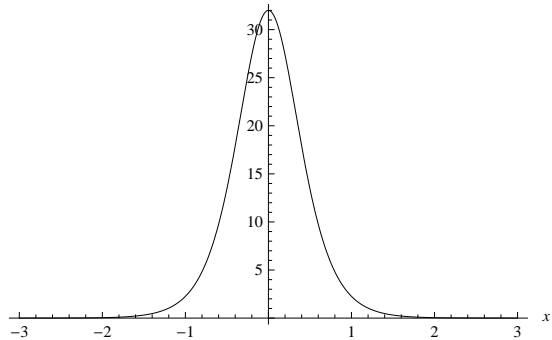}
\caption{Plot of P\"oschl-Teller potential $V$ of Eq.\ (\ref{eq: potential Poschl-Teller}) corresponding to $\omega_V^+$, where we take $D=5$, $\Lambda=1$, and $l=3$.} 
\end{figure}

\begin{figure}[th]
\label{figure2}
\includegraphics[scale=.85,clip=true]{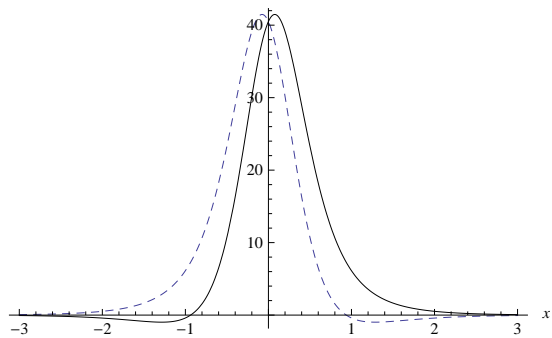}
\caption{ Plot of Morse potentials $V_+$ (solid line) and $V_-$ (dashed line) of Eq.\ (\ref{eq: Morse potential}), where we take $\hat{m}=0$, $D=5$, $\Lambda=1$, and $l=3$.} 
\end{figure}

\section{Concluding remarks}
\label{sect 4}

The $D$-di\-men\-sion\-al Nariai spacetime studied in Section \ref{sect 3} is uncharged. Notice that there is a charged generalization of the $D$-di\-men\-sion\-al Nariai spacetime, only we need to replace the quantity $\sigma$ of formula (\ref{eq: sigma Nariai}) by \cite{Kodama:2003kk}, \cite{b: Nariai solution}
\begin{equation} \label{eq: sigma charged}
 \sigma_Q = (D-1) \Lambda - \frac{(D-3)^2 Q^2}{a^{2 (D-2)}},
\end{equation} 
and the parameter $a$ of expression (\ref{eq: a Nariai}) is replaced in the charged case by $a_Q$ which is a solution to the equation
\begin{equation} \label{eq: a charged}
 \frac{(D-3)}{a_Q^2} = (D-1)\Lambda + \frac{(D-3)Q^2}{a_Q^{2 (D-2)}},
\end{equation} 
where $Q$ is related to the electric charge of the spacetime. 

Supported in our mathematical analysis of the problem for the uncharged Nariai spacetime, we assert that in the charged $D$-di\-men\-sion\-al Nariai spacetime the QNF of the uncharged Dirac field are determined by formulae (\ref{eq: QN frequencies Nariai}), only we must replace in these formulae the values of the quantities $\sigma$ and $a$ by $\sigma_Q$ and $a_Q$, respectively. 

For identical values of the parameters $D$, $\Lambda$, and $n$, Morse potentials (\ref{eq: Morse potential}) and P\"oschl-Teller potentials (\ref{eq: potential Poschl-Teller}) have QNF with identical imaginary parts, that is with identical decay times, even when for identical values of $D$, $\Lambda$, and $l$ the height of P\"oschl-Teller potential is smaller than the height of Morse potentials (see Figures 1 and 2). We believe that to find the source of this coincidence is an interesting question. 

Also, in Ref.\ \cite{Beyer:1998nu} was shown that for sufficiently late times the radial functions of P\"oschl-Teller potential (\ref{eq: potential Poschl-Teller}) form a complete basis. Owing to similarity of the plots for both potentials (see again Figures 1 and 2), to study if a similar result is valid for Morse potential (\ref{eq: Morse potential}) deserves a detailed investigation. 

As we previously comment a similar reduction to that of Section \ref{sect 2} works for charged Dirac fields propagating in the $D$-di\-men\-sion\-al charged spherically symmetric spacetimes \cite{Gibbons:1993hg}. We believe that a good exercise is to calculate the QNF of the charged Dirac field propagating in the $D$-di\-men\-sion\-al charged Nariai spacetime to extend the results of Refs.\ \cite{LopezOrtega:2007vu}, \cite{Vanzo:2004fy} and the previous section.

As we observe in Introduction section, the results on the separability of the Dirac equation in the four-di\-men\-sion\-al Kerr black hole generalize to some $D$-di\-men\-sion\-al rotating black holes \cite{Unruh:1974bw}, \cite{Chandrasekhar:1976ap}. We believe that the extension of the results obtained in these references to the metrics of Plebanski-Demianski-Klemm type \cite{Klemm:1998kd} deserve a detailed analysis. Furthermore the study of the separability properties of the equations of motion for gravitational and electromagnetic perturbations in the $D$-di\-men\-sion\-al Myers-Perry metrics of Ref.\ \cite{Myers:1986un} is a relevant problem.

\section{Acknowledgements}

I thank Dr.\ C.\ E.\ Mora Ley, Dr.\ R.\ Garc\'{\i}a Salcedo,  Dr.\ O.\ Pedraza Ortega and A.\ Tellez Felipe for their interest in this paper. This work was supported by CONACYT M\'exico, SNI M\'exico, EDI-IPN, COFAA-IPN, and Research Project SIP-20090952.

\end{document}